\documentclass[longauth]{aa} 
\usepackage{graphicx}
\usepackage{epstopdf}
\usepackage{txfonts}
\usepackage{float}

\begin{document}
%

\title{Constraints on high-energy neutrino emission from SN 2008D}

\author{
IceCube Collaboration:
R.~Abbasi\inst{1},
Y.~Abdou\inst{2},
T.~Abu-Zayyad\inst{3},
J.~Adams\inst{4},
J.~A.~Aguilar\inst{1},
M.~Ahlers\inst{5},
K.~Andeen\inst{1},
J.~Auffenberg\inst{6},
X.~Bai\inst{7},
M.~Baker\inst{1}
% ...
}
\institute{Dept.~of Physics, University of Wisconsin, Madison, WI 53706, USA\and
Dept.~of Subatomic and Radiation Physics, University of Gent, B-9000 Gent, Belgium\and
Dept.~of Physics, University of Wisconsin, River Falls, WI 54022, USA\and
Dept.~of Physics and Astronomy, University of Canterbury, Private Bag 4800, Christchurch, New Zealand\and
Dept.~of Physics, University of Oxford, 1 Keble Road, Oxford OX1 3NP, UK\and
Dept.~of Physics, University of Wuppertal, D-42119 Wuppertal, Germany\and
Bartol Research Institute and Department of Physics and Astronomy, University of Delaware, Newark, DE 19716, USA
% ...
}

\date{Received September 16, 2010; accepted December 9, 2010}

\abstract{}{}{}{}{} 
 
  \abstract
   {SN 2008D, a core collapse supernova at a distance of 27 Mpc, was serendipitously discovered by the \textit{Swift} satellite through an associated X-ray flash. Core collapse supernovae have been observed in association with long gamma-ray bursts and X-ray flashes and a physical connection is widely assumed. This connection could imply that some core collapse supernovae possess mildly relativistic jets in which high-energy neutrinos are produced through proton-proton collisions. The predicted neutrino spectra would be detectable by Cherenkov neutrino detectors like IceCube. }
   {A search for a neutrino signal in temporal and spatial correlation with the observed X-ray flash of SN 2008D was conducted using data taken in 2007-2008 with 22 strings of the IceCube detector.  }
   {Events were selected based on a boosted decision tree classifier trained with simulated signal and experimental background data. The classifier was optimized to the position and a ``soft jet'' neutrino spectrum assumed for SN 2008D. Using three search windows placed around the X-ray peak, emission time scales from $100 - 10000$ s were probed.}
   {No events passing the cuts were observed in agreement with the signal expectation of 0.13 events. Upper limits on the muon neutrino flux from core collapse supernovae were derived for different emission time scales and the principal model parameters were constrained. 
	}
   {While no meaningful limits can be given in the case of an isotropic neutrino emission, the parameter space for a jetted emission can be constrained. Future analyses with the full 86 string IceCube detector could detect up to $\sim$100 events for a core-collapse supernova at 10 Mpc according to the soft jet model. }
   \keywords{core collapse supernovae -- 
		SN 2008D --
                cosmic neutrinos --
                SN-GRB connection --
		high-energy neutrinos
               }
\authorrunning{N. Kemming for the IceCube Collaboration}
\titlerunning{Constraints on high-energy neutrino emission from SN 2008D}
   \maketitle
%

\section{Introduction}

Observations in recent years have given rise to the idea that core collapse supernovae (SNe) and long duration gamma-ray bursts (GRB) have a common origin or may even be two different aspects of the same physical phenomenon, the death of a massive star with $M >8\,M_{\odot}$ (for a review, see Woosley, Bloom \cite{woosley}). Like GRBs, SNe could produce jets, though less energetic and collimated and possibly ``choked'' within the stellar envelope. Observed associations of supernovae with XRFs, short X-ray flashes with similar characteristics to long GRBs, suggest including XRFs in the SN-GRB connection as well. Although XRFs are considered a separate observational category from GRBs, a common origin and a continuous sequence connecting them have been suggested (Lamb et al. \cite{lamb}, Yamazakia et al. \cite{yamazakia}). XRF could be long GRBs with very weak jets or simply long GRBs observed off-axis. Several XRFs or or long duration, soft-spectrum GRBs have been observed in coincidence with core collapse SNe thus far: SN 1998bw (Galama et al. \cite{galama}), SN 2003lw (Malesani et al. \cite{malesani}), SN 2003dh (Hjorth et al. \cite{hjorth}), SN 2006aj (Pian et al. \cite{pian}), and of course SN 2008D (Soderberg et al. \cite{soderberg}, Modjaz et al. \cite{modjaz}, Mazzali et al. \cite{mazzali}). For SN 2007gr (Paragi et al. \cite{paragi}) and SN 2009bb (Soderberg et al. \cite{soderberg2009bb}), two core collapse SNe not associated with an XRF or GRB, recent radio observations provide strong evidence for jets with bulk Lorentz factors of $\,\Gamma > 1$. If some core collapse SNe indeed form such "soft'' jets, protons accelerated within the jet could produce TeV neutrinos in collisions with protons of the stellar envelope (Razzaque et al. \cite{rmw}, Ando \& Beacom \cite{ab}). The soft jet scenario for core collapse SNe can be probed with high-energy neutrinos even if the predicted jets stall within the stellar envelope and are undetectable in electromagnetic observations.

On January 9, 2008, the X-ray telescope aboard the SWIFT satellite serendipitously discovered a bright X-ray flash during a pre-scheduled
observation of NGC 2770. Optical follow-up observations were immediately triggered and recorded the optical signature of SN 2008D, a core collapse supernova of type Ib  at right ascension $\alpha = 09\,\mathrm{h}\,\,09\,\mathrm{m}\,\,30.70\,\mathrm{s}\,$ and declination $\delta = 33^{\circ}\,08'\,19.1"\,$ (Soderberg et al. 2008). SN 2008D offers a realistic chance to detect high-energy supernova neutrinos for the first time since the observed X-ray peak provides the most precise timing information ever available to such a search. Whether or not the existence of jets in aspherical explosions is evidenced in the spectroscopic data for SN 2008D remains highly debated. While Soderberg et al. (\cite{soderberg}) ``firmly rule out'' any asphericity and Chevalier and Fransson (\cite{chevalierfransson}) speak of a purely spherical shock-breakout emission, Mazzali et al. (\cite{mazzali}) and Tanaka et al. (\cite{tanaka}) find evidence that SN 2008D possessed jets which have been observed significantly off-axis.

The IceCube neutrino detector, currently under construction
at the South Pole and scheduled for completion in 2011, is capable
of detecting high-energy neutrinos ($E_{\nu} \gtrsim 100\,\mathrm{GeV}$) of
cosmic origin by measuring the Cherenkov light emitted by
secondary muons with an array of Digital Optical Modules
(DOMs) positioned in the transparent deep ice along vertical
strings (J. Ahrens et al. \cite{icecube}). The full detector will
comprise 4,800 DOMs deployed on 80 strings between 1.5 and 2.5 km
deep within the ice, a surface array (IceTop) for observing
extensive air showers of cosmic rays, and an additional
dense subarray (DeepCore) in the detector center
for enhanced low-energy sensitivity. Each DOM consists
of a 25 cm diameter Hamamatsu photo-multiplier
tube (PMT, see Abbasi et al. \cite{pmt}), electronics for waveform digitization (Abbasi et al. \cite{daq}), 
high voltage generation, and a spherical, pressure-resistant glass housing.
The DOMs detect Cherenkov photons emitted by relativistic
charged particles passing through the ice. In
particular, the directions of muons (either from cosmic
ray showers above the surface or neutrino interactions
within the ice or bedrock) can be well reconstructed from
the track-like pattern and timing of hit DOMs. Identification
of neutrino-induced muon events in IceCube has
been demonstrated in Achterberg et al. (\cite{ic9}) using atmospheric
neutrinos as a calibration tool. 
Sources in the northern sky, like SN
2008D, can be observed with very little background since contamination by atmospheric muon tracks is eliminated by the shielding effect of the Earth. When SN 2008D was discovered, the installation of IceCube
was about one quarter completed and the detector was taking
data with 22 strings.

As shown above, a search for cosmic neutrinos from core collapse SNe is motivated by both observational evidence and theoretical predictions. While analyses using catalogs of SNe/GRBs with timing uncertainties $\sim$1 d as the signal hypothesis have been performed on archived AMANDA/IceCube data (see Lennarz  \cite{lennarz} for SNe and Abbasi et al. \cite{grbapj} for GRBs), the unprecedentedly precise timing information available for SN 2008D suggests a designated study of this event. While electromagnetic observations provide no conclusive evidence for the existence of highly relativistic jets, soft, hidden jets could be revealed by high energy neutrinos, assuming sufficient alignment with the line of sight.

The paper is organized as follows: Section 2 discusses the assumed model for neutrino production. Section 3 describes the experimental and simulated data used for the analyis. The selection criteria used to separate signal events from background are detailed in Section 4. Section 5 presents the results of the search and constraints derived therefrom. Finally, the analysis is summarized in Section 6.


\section{Model neutrino spectrum}
\label{SecModel}
A model for the emission of high-energy neutrinos in jets formed by core collapse supernovae has been proposed by Razzaque, Meszaros, and Waxman (\cite{rmw}) and further elaborated by Ando and Beacom (\cite{ab}). This model will be referred to as ``soft jet model'' in the following. A brief summary of the physical motivation and a derivation of its analytical form shall be presented. 

\begin{figure}
   \centering
	\caption{Assumed $E^2$-weighted muon neutrino and antineutrino spectrum of SN 2008D according to the soft jet model. 
	For comparison, the atmospheric muon neutrino flux is shown for a $100\,\mathrm{s}$ time window and a circular aperture with opening angle $\omega = 10^{\circ}$.
              }
   \includegraphics[width=9cm]{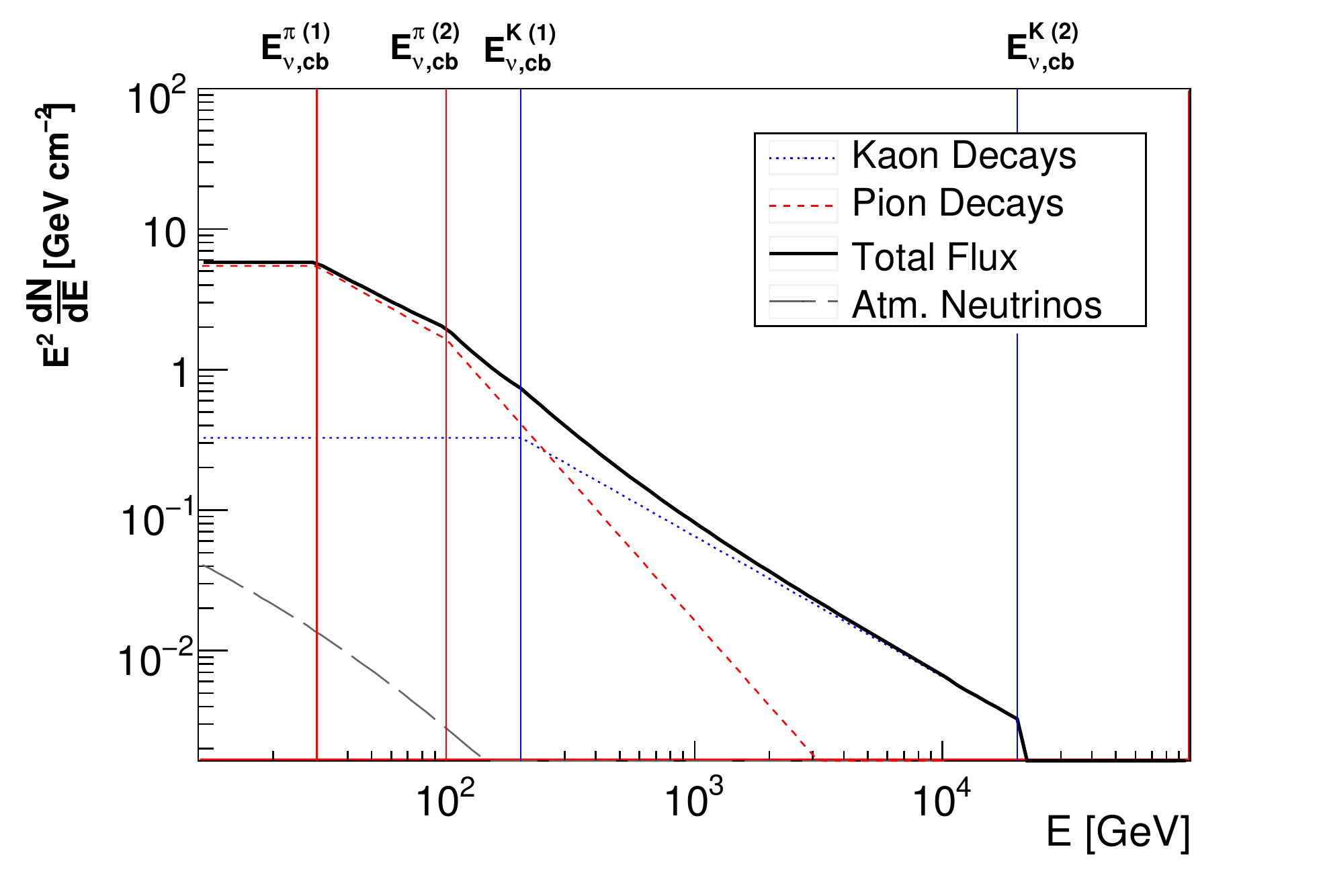}
      
         \label{FigSlowJetModel}
   \end{figure}

The soft jet model assumes the collapse of a massive star $M_{\star}\gtrsim 8\,M_{\odot}$ with subsequent formation of a neutron star or black hole, rotating sufficiently to power jets with bulk Lorentz factors of $\Gamma_b \sim 1-10$ and opening angles $\theta_j \approx 1/\Gamma_b = 5^{\circ}-50^{\circ}$. Such ``soft'' jets, too weak to penetrate the stellar envelope, would not be observable in the electromagnetic spectrum. The rebounding core collapse is assumed to deposit $E_{j}\sim 3\times 10^{51}\,\mathrm{erg}$ of kinetic energy in the material ejected in the jets -- values of up to $E_j = 6\times 10^{51}\,\mathrm{erg}$ have been suggested for SN 2008D by Mazzali et al. (\cite{mazzali}). Protons are Fermi accelerated to a $E_{p}^{-2}$-spectrum and produce muon neutrinos through the decay of charged pions and kaons formed in proton-proton collisions. 
The neutrino spectrum, shown in Fig. \ref{FigSlowJetModel}, follows the primary proton spectrum at low energies and steepens at four break energies above which pions (kaons) lose a significant fraction of their energy in hadronic and radiative cooling reactions, before decaying into neutrinos. These break energies are distinct for pions and kaons and exihibit a sensitive dependence on the jet parameters (see Table \ref{TabModelParams}). Using the notation of Ando and Beacom, the spectrum can be written as:
\begin{equation}
\displaystyle \Phi_{\nu} (E_{\nu}) =\sum\limits_{i\,=\,\pi,K} \eta_i \times
\left\lbrace
\begin{array}{ll}
\,\,\displaystyle E_{\nu}^{-2}                   &\,\, E_{p,min} \leq E_{\nu} < E_{\nu,cb}^{i\,(1)} \\
\,\,E_{\nu}^{-3}\,E_{\nu,cb}^{i\,(1)}\                &\,\, E_{\nu,cb}^{i\,(1)} \leq E_{\nu} < E_{\nu,cb}^{i\,(2)} \\
\,\,E_{\nu}^{-4}\,E_{\nu,cb}^{i\,(1)}\,E_{\nu,cb}^{i\,(2)} &\,\, E_{\nu,cb}^{i\,(2)} \leq E_{\nu} \leq E^{i}_{\nu,max} \\
\end{array}
\right.
\label{EqSlowJetModel}
\end{equation}
where
\begin{equation}
\eta_i = \frac{\displaystyle \langle n \rangle_i\,B_i\,E_j}{\displaystyle 8\,\left(2\,\pi\,\theta_j^2\,d^2\right)\,\ln{\left( E_{p,max}/E_{p,min}\right) }}
\label{EqNormalization}
\end{equation}
With the exception of the distance $d$, we assume the same parameters for SN 2008D that are quoted in Ando \& Beacom (\cite{ab}). A summary is given in Table \ref{TabModelParams}.

%
%

\begin{table}
\caption{Parameters of the soft jet model used in this analysis.}
\label{TabModelParams}
\begin{tabular}{p{0.12\linewidth}p{0.35\linewidth}p{0.19\linewidth}p{0.16\linewidth}}
\hline\noalign{\smallskip}
Parameter & Description & Default Value & Parameter Dependence \\
\noalign{\smallskip}\hline\noalign{\smallskip}
$E_j \,$ & Total kinetic energy of ejected material & $10^{51.5} \mathrm{erg}$ &  - \\
\noalign{\smallskip}
$\Gamma_b \,$ & Bulk Lorentz factor of the jet & $3 \,$ & - \\
\noalign{\smallskip}
$d \,$ & Distance of SN 2008D & $27\,\mathrm{Mpc} \,$ &  - \\ 
\noalign{\smallskip}
$B_{\pi}$ $\left(B_{K}\right)$ & Branching ratio for \newline $\pi^{\pm} \rightarrow \mu^{\pm}\nu_{\mu}$  ($K^{\pm} \rightarrow \mu^{\pm}\nu_{\mu}$)& $1$  ($\,0.63\,$) & - \\
\noalign{\smallskip}
$\langle n \rangle_{\pi} \,$ $\left(\,\langle n \rangle_{K}\,\right)$ & Pion (kaon) multiplicity in $pp$ collisions & $1 \,$ ($\,0.1\,$) & - \\
\noalign{\smallskip}
$E_{p,min}$ & mininmum proton energy & $10\,\mathrm{GeV}$  & - \\
\noalign{\smallskip}
$E_{p,max}\,$ & maximum proton energy & $7\times 10^4\,\mathrm{GeV}$ & - \\
\noalign{\smallskip}
$E_{\nu,cb}^{\pi(1)}$ \newline $\left( E_{\nu,cb}^{K(1)}\right) $ & hadronic cooling break energy for pions (kaons) & $30\, \mathrm{GeV}$\newline  ($200\,\mathrm{GeV} \,$) & $\propto E_j^{-1}\,\Gamma_b^7\,\theta_j^2\,$ \newline
$\,\,\,\,\,\,\approx E_j^{-1}\,\Gamma_b^5\,$ \\
\noalign{\smallskip}
$E_{\nu,cb}^{\pi(2)}$ \newline $\left( E_{\nu,cb}^{K(2)}\right) $ & radiative cooling break energy for pions (kaons) & $100 \,\mathrm{GeV}$ \newline  ($20000 \,\mathrm{GeV}$) & $\propto \Gamma_b$ \\ 
\noalign{\smallskip}
$E^{\pi}_{\nu,max}$ \newline $\left(E^{K}_{\nu,max}\right) \,$ & maximum neutrino energy from pion (kaon) decay & $10500\,\mathrm{GeV}$ \newline($21000\,\mathrm{GeV}$) &  $\propto \Gamma_b$ \\
\noalign{\smallskip}
\hline
            
\end{tabular}
\end{table}

An optimistic extension of this model proposed by Koers and Wijers (\cite{koers}) predicts that mesons are again Fermi-accelerated after production. This re-acceleration gives rise to a simple $E^{-\gamma}$ neutrino spectrum with $\gamma = 2.0,\dots, 2.6$ extending to maximum energies of $E_{\nu}\sim10\,\mathrm{PeV}$ where radiative cooling processes lead to a steepening and eventual cutoff of the neutrino spectrum. The details of this high-energy cutoff are negligible in the context of this analysis, where neutrinos with energies of 100 GeV - 10 TeV are expected to yield the dominant contribution of the signal expectation.

Neutrinos are expected to be emitted in alignment with the jets.
Their energy range is set by the maximum proton energy and reaches far into the sensitive range of the IceCube detector ($E_{\nu}\gtrsim 100\,\mathrm{GeV}$). In order to detect these neutrinos, the jet must be pointing towards Earth (e.g. 5\% chance for a jet with an opening half angle of $17^{\circ}$). Due to the unknown jet pointing, however, no  constraints can be placed on the model in the case of a non-detection. To do so with a confidence level of e.g. 90\% would require a large sample of $\sim$200 nearby supernovae. In contrast, a positive detection would not only indicate the jet's direction, but also yield constraints on the soft jet model -- constraints entirely independent of observations in the electromagnetic spectrum. If, in addition, a resolved neutrino spectrum could be recorded with future neutrino detectors, the observation of spectral breaks and a spectral cutoff would place strong constraints on the physical parameters of the supernova jet.

\section{Data and Simulation}
The analysis uses experimental data to determine the expected number of background events for a particular search window. The signal expectations as well as the characterictics of the signal are derived from simulations. Raw data consists of time series of photon detections (henceforth ``hits'') for each triggered DOM. From these hit patterns, track reconstruction algorithms derive the muon's direction, measured in zenith $\theta$ and azimuth $\phi$ in a fixed detector coordinate system where muons travelling upwards in the ice have $\theta>90^{\circ}$ and downgoing tracks have $\theta < 90^{\circ}$. The absolute time of an event is determined by a GPS clock with a precision of better than $200\,\mathrm{ns}$, which is more than sufficient for this analysis.

\subsection{Background data}
At trigger level (detailed in Sec. \ref{SecTrigger} below), IceCube data is dominated by the \textit{reducible} background of atmospheric muons, falsely reconstructed as upgoing, i.e. having passed through the Earth. A comparison of experimental data and simulated muons from cosmic ray showers shows good agreement (see Fig. \ref{FigCutParams}). In addition, background data contains an \textit{irreducible} background of muons produced by atmospheric neutrinos from the northern hemisphere, at a rate lower by a factor of $10^5$. At the final cut levels of this analysis (see Tab. \ref{TabWindows}), data consists of approximately equal contributions of reducible and irreducible background events.

The data sample used to measure and characterize background was taken by IceCube in the
22 string configuration over 275.72 days of detector live time between May 2007 and March
2008. The sample is identical to the one used in the first IceCube search for neutrino point sources (R. Abbasi et al. \cite{pointsource}).
On the day of SN 2008D, IceCube was taking data continuously in a time range of $[-9.5\,\mathrm{h},\,+1.8\,\mathrm{h}]$ around the observed X-ray flash.
To prevent a bias in the cut optimization, this data was kept ``blind'', i.e. excluded from the development and testing of selection criteria, and only ``unblinded'' in the final step of the analysis.

\subsection{Signal Simulation}
To quantify and characterize the expected signal, extensive simulations of the complex Earth--ice--detector system were conducted. IceCube simulation generates primary neutrinos at the surface of the Earth and propagates them through the Earth, tracking charged and neutral current interactions, and recording all secondary particles which can reach the detector (see Kowalski et al. \cite{anis}).
All secondary muons are then passed to the muon propagation software (see Chirkin \& Rhode \cite{mmc}) which simulates their random energy loss and the emission of Cherenkov photons. Finally, the propagation of photons is simulated accounting for absorption and scattering according to a depth dependent ice model (see Lundberg et al. \cite{photonics}). In the last step, the photomultiplier response, readout, and local as well as global triggers are simulated yielding time series of photon hits which are subsequently passed through the same processing pipeline as experimental data.

\subsection{Triggering and data processing}
\label{SecTrigger}
The IceCube trigger system only reads out a photon hit at a specific optical module if a neighboring module on the same string is also hit within $1\,\mu\mathrm{s}$ (local coincidence). To initiate the event read-out, the global trigger of IceCube 22 required 8 such local coincidences within a $5\,\mu$s time window. This requirement lead to trigger rates of $\sim$550 Hz, dominated by atmospheric muon events. Data contamination was immediately reduced to $\sim$25 Hz by first-guess reconstructions running online at the South Pole, which fit a simple track hypothesis to each event and reject downgoing tracks in real time (Ahrens et al. \cite{trackreco}). Events passing this online muon filter are transferred to the North, where extensive likelihood track reconstructions are performed.
For a given hit pattern and a first guess track hypothesis, the likelihood function is calculated as the product of the probabilities for each hit time to occur under the given track hypothesis. The likelihood reconstruction algorithm then iteratively searches for the track which maximizes the value of this likelihood function (Ahrens et al. \cite{trackreco}). For the final fit result, the optimization sofware computes quality parameters which can be used for event selection.

\section{Event selection}
The background event rate is further diminished to $\sim$3 Hz through another cut on the more precise track direction from the likelihood track reconstruction selecting events with $\theta > 80^{\circ}$. For this analysis, events outside a circular signal region ($10^{\circ}$ opening angle) around the position of SN 2008D were removed from the dataset to obtain a manageably sized sample. At this filtering level, the background rate is 0.03 Hz and 0.26 signal events are expected for SN 2008D according to the soft jet model.
 
\subsection{Quality Cuts}
Specific cuts tailored to the simulated properties of SN 2008D were based on the following eight quality parameters:
\\ \\
\begin{tabular}{cp{0.8\linewidth}}
	    $N_{dir,E}$                            &
Number of direct hits, i.e. photon hits detected within a $[-15\,\mathrm{ns},\,+250\,\mathrm{ns}]$ time window of the arrival time predicted for unscattered Cherenkov emission under the track hypothesis \\ \smallskip
 	    $S_{all}$                              & 
Smoothness of hit distribution. $S_{all}=0$ indicates a homogeneous energy deposition along the track \\ \smallskip
            $\theta_{min}$                      & 
Minimum zenith when the 1$^{\mathrm{st}}$ and 2$^{\mathrm{nd}}$ half of the photon hits (ordered in time) are reconstructed as separate tracks \\ \smallskip
	    $\sigma_{p}$                           & 
Estimator for the uncertainty of the reconstructed track direction (quadratic average of the minor and major axis of the $1\,\sigma$ error ellipse)\\  \smallskip
            \begin{math} \mathcal{L}_{R}\end{math} & 
Value of the negative log-likelihood for the reconstructed track divided by the number of
degrees of freedom in the fit (number of hit optical modules minus number of fit parameters)  \\ \smallskip
            $R_{B}$                                & 
Ratio of the log-likelihoods with and without a Bayesian prior that favors a downgoing track hypothesis\\ \smallskip
            $R_{U}$                                & 
Ratio of the log-likelihoods with and without seeding the reconstruction with the inverse track direction \\ 
\end{tabular}
\\
\noindent In conjunction with the selection of upgoing tracks, the reduced log-likelihood $L_R$ has proven to be an efficient variable for separating
upgoing atmospheric neutrinos from misreconstructed downgoing atmospheric muons. It exploits the fact that for a light pattern originating from a downgoing muon the incorrect upgoing track hypothesis yields rather low \textit{absolute} likelihood values. In addition, the likelihood ratios $R_U$ and $R_B$ allow for a veto on events for which inverting the track hypothesis leads to a significant \textit{relative} enhancement in the likelihood value.
 
\begin{figure}
   \centering
\caption{Normalized histograms of cut parameters used for event selection (top) and resulting distribution of boosted decision tree classifier values (bottom). Dashed line indicates experimental background data, dotted line marks background simulation, solid line represents simulated signal.}
   \includegraphics[width=9.1cm]{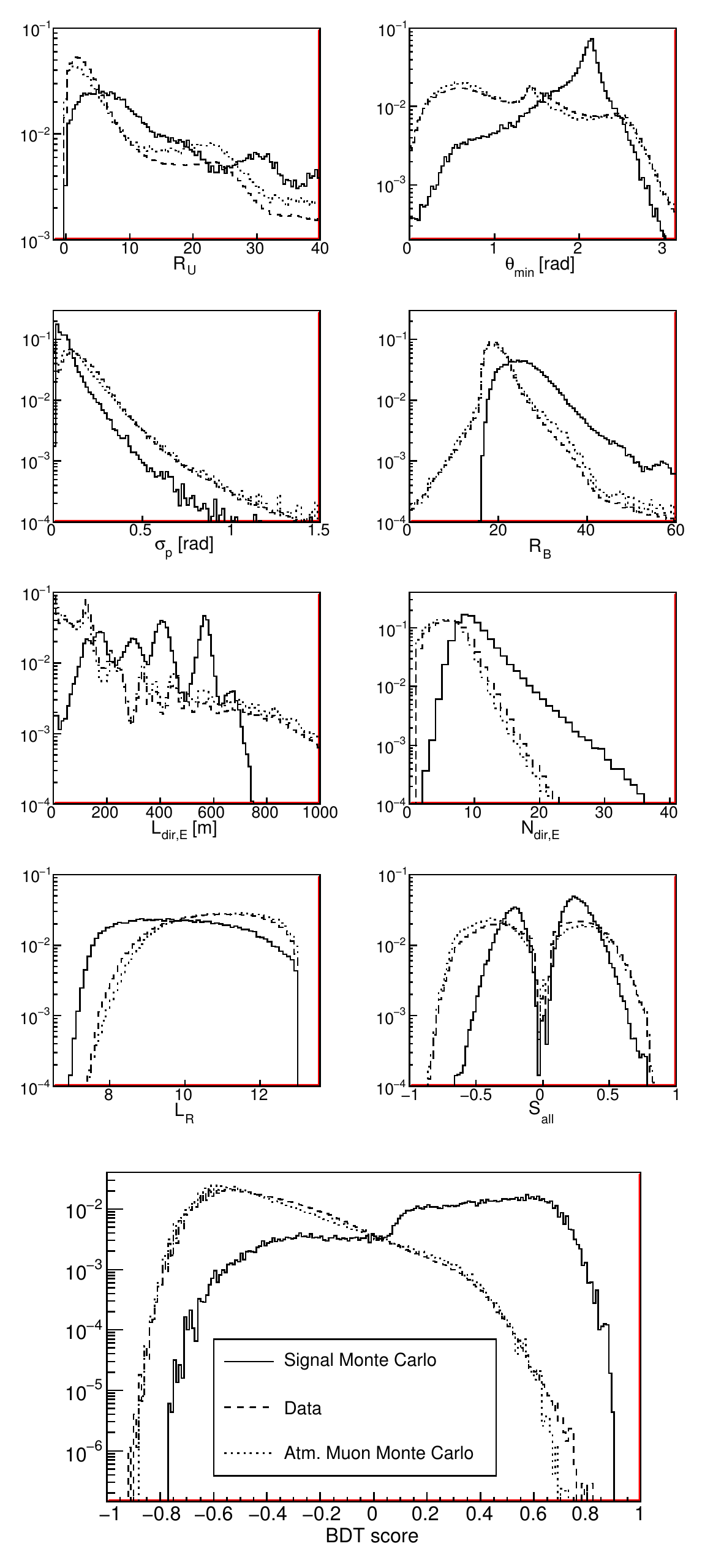}
	
              \label{FigCutParams}%
    \end{figure}

Histograms of all selection parameters are shown in Fig. \ref{FigCutParams} for background data, background simulation, and simulated signal events. To combine all eight parameters efficiently, they were incorporated into a boosted decision tree (BDT) classifier (see e.g. Yang et al. \cite{bdt} and references therein). The BDT method classifies an event by passing it through a tree structure of binary splits which effectively breaks up the parameter space into a number of signal or background-like hypercubes. The classifier is first trained with background data and simulated signal and then evaluated with independent datasets. The resulting distribution of classifier scores $\mathcal{K}$ for experimental data and simulated signal is shown at the bottom of Fig. \ref{FigCutParams}. The classifier allows for a simple one-dimensional cut on the classification score. Extensive tests were conducted to assure a stable response and to estimate the uncertainty of the classification. This uncertainty was estimated by comparing the classification efficiencies for several independent experimental data and simulated signal samples. Variations in the classifier response proved to be negligible compared to statistical uncertainties.

\subsection{Search Windows}
\begin{figure}
   \centering
\caption{Effective areas for a neutrino spectrum obeying the soft jet model. Each line respresents one of the final search windows used in this analysis. Inset: Cumulative point spread function for the direction in which SN 2008D was observed. }
   \includegraphics[width=\linewidth]{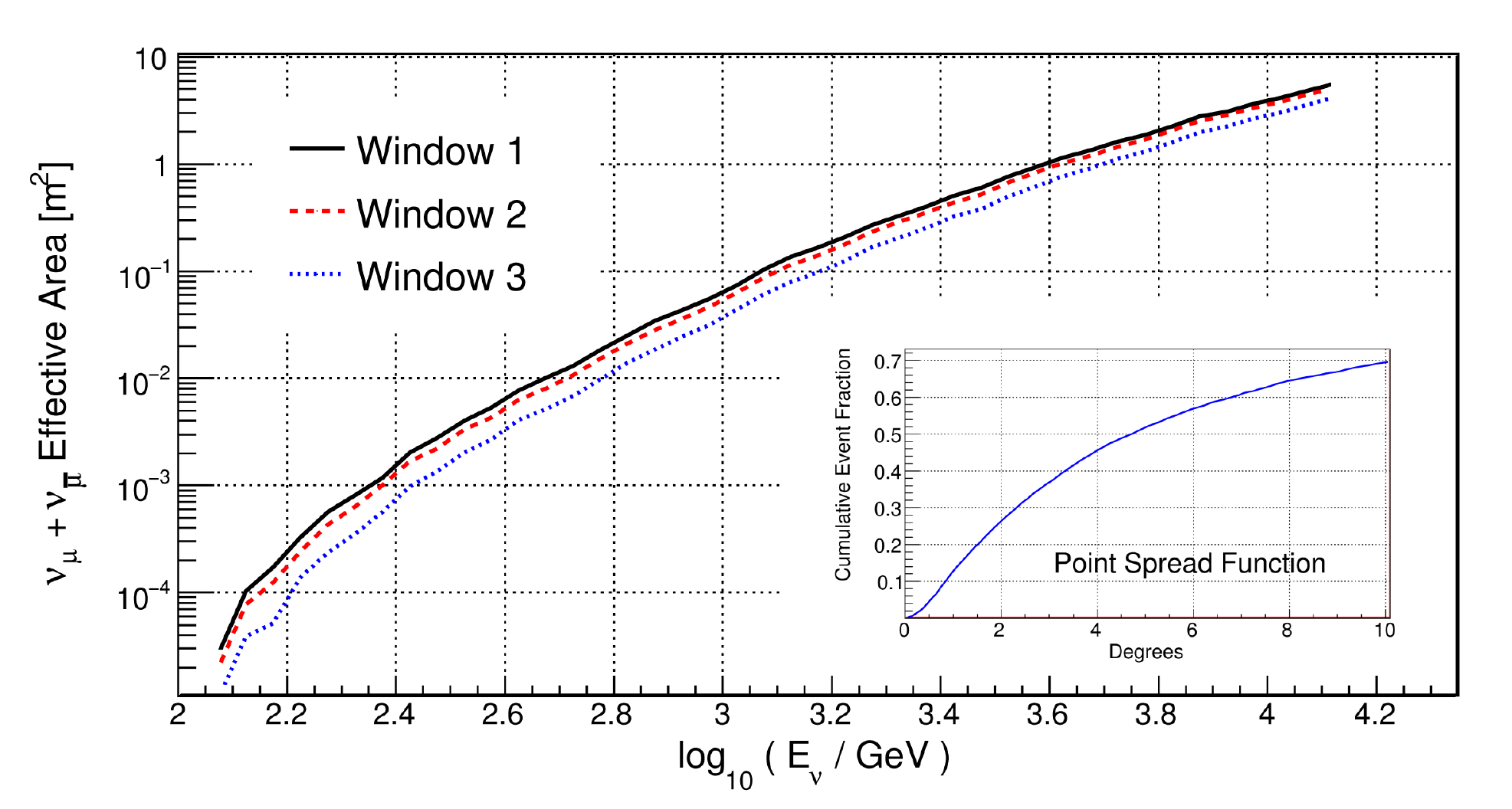}\\	
              \label{FigAeff}
    \end{figure}

The search for neutrinos in the on-time data from January 9, 2008 was conducted using three search windows of different durations, apertures, and selection cuts. A circular aperture was used in all cases. Since the soft jet model does not explicitly predict the time profile of the neutrino emission, search windows with durations of $100\,\mathrm{s}$, $1000\,\mathrm{s}$, and $10000\,\mathrm{s}$ were chosen to cover a large range of emisssion time scales. The corresponding opening angle and quality cuts for each search window were determined by optimizing the model discovery factor $M$ according to Hill et al. (\cite{hill}). For this purpose, a Poisson distribution with mean $\,b+s\,$ is randomly sampled, where $b$ and $s$ represent the expected background and signal, respectively. For each drawn number of observed events $n_{\mathrm{obs}}$ the lower limit on the signal contribution is computed using the Feldman\&Cousins algorithm (Feldman\&Cousins \cite{feldmancousins}). The signal expectation is increased $s\rightarrow s^{\star}$ until 50\% of the trials yield a discovery, that is, a lower limit on the signal $s$ greater than zero. When this criterion is met, the model discovery factor is given by
\begin{equation}
 M=\frac{\,\,s^{\star}}{s}
\end{equation}
For each window, the BDT cut $\mathcal{K}$ and the opening angle $\omega$ yielding the minimal value of $M$ were determined numerically. Lower limits according to the Feldman\&Cousins ordering scheme were required to have a significance of $5\,\sigma$. The choices of cuts for the three search windows which yielded minimal model discovery factors are summarized in Table \ref{TabWindows}. The resulting effective areas for a neutrino spectrum obeying the soft jet model are shown in Fig. \ref{FigAeff}.

   \begin{table}
	 \caption{Windows used to search for neutrinos in correlation with SN 2008D. For each time scale, quality and angular cuts were optimized to yield a maximum model discovery potential.}\label{TabWindows}
         \begin{tabular}{p{0.15\linewidth}p{0.12\linewidth}p{0.25\linewidth}p{0.12\linewidth}p{0.13\linewidth}}
            \hline
            \noalign{\smallskip}
              &  Duration \newline $\Delta t$ & Centering\newline wrt. X-ray peak & Aperture\newline $\omega$ & BDT cut \newline $\mathcal{K}$ \\
            \noalign{\smallskip}
            \hline
            \noalign{\smallskip}
        	Window 1 & $100$ s & $-70\,\mathrm{s}$, $\,\,+30\,\mathrm{s}$ & $6.2^{\circ}$ & $0.390$ \\
Window 2 & $1000$ s & $-500\,\mathrm{s}$, $\,+500\,\mathrm{s}$ &  $2.6^{\circ}$ & $0.464$\\
Window 3 & $10000$ s & $-7000\,\mathrm{s}$, $+3000\,\mathrm{s}$ &  $1.5^{\circ}$ & $0.580$\\
	\noalign{\smallskip}
            \hline
         \end{tabular}
   \end{table}

With these choices, two observed events would constitute a $5\,\sigma$ discovery in any of the windows taken by itself. The significances for the complete measurement consisting of three search windows were determined in a simulation study with $10^{10}$ trials. For each possible observation of $n_1,\,n_2,\, n_3$ events in window 1, 2, 3, the p-value was calculated as the fraction of equally or less likely observations.

\section{Results}

\subsection{Unblinding}
No events passing the cuts were found in the experimental data. As shown in Table \ref{TabResults}, this result is consistent with expectations, even more so if we account for the $\sim$5\% probability of a jet with opening half angle $\sim$17$^{\circ}$ pointing towards Earth.

\begin{table}
\centering
\caption{Summary of the unblinding results and comparison with expectations.}
\label{TabResults}
\begin{tabular}{lcccc}
	\hline\noalign{\smallskip}
	& & Window 1 & Window 2  &  Window 3  \\ \noalign{\smallskip}\hline\noalign{\smallskip}
	Observed Events &  & $n_1 = 0$ & $n_2 = 0$ & $n_3 = 0$ \\
	Expected Events & & & \\
	$\quad$Signal $s$ & & 0.13 & 0.060 & 0.020 \\
	$\quad$Background $b$ & & $3.67 \times 10^{-4}$ &  $5.52 \times 10^{-4}$& $5.55 \times 10^{-4}$ \\
	\noalign{\smallskip}\hline
\end{tabular}
\end{table}

\subsection{Limits on the Soft Jet Model}
In the absence of more precise theoretical predictions on the time profile of the emission, quoting limits for particular time scales is the only viable way to constrain the soft jet model. Since $n_1 =n_2=n_3 =0$ and $b_1 \approx b_2 \approx b_3$, the signal upper limits $\bar{s}_i$ are identical for all three search windows to the fourth significant digit: $\,\bar{s}_1=\bar{s}_2=\bar{s}_3=\bar{s}=2.44\, $ (at 90\% CL). The upper limit $\,\bar{\Phi}_{\nu}^{(90)}\,$ on the neutrino flux in terms of the expected flux $\Phi_{\nu}$ is given by the ratio of the signal upper limit $\bar{s}$ to the signal expectation $s$:
\begin{equation}
	\frac{\,\,\,\bar{\Phi}_{\nu}^{(90)}}{\Phi_{\nu}} = \frac{\bar{s}}{s} 
\label{EqUpperLimit}
\end{equation}
  Due to the different signal expectations in each window, the flux upper limits depend on the assumed emission time scale $\tau_e$. Therefore, we quote the limits on the soft jet model for canonical parameters (Tab. \ref{TabModelParams}) separately for each emission time scale $\tau_e$ and at a reference energy of $E_{\nu}=100\,\mathrm{GeV}$:
\begin{equation}
	\left[ \frac{\bar{\Phi}_{\nu}^{(90)}\left( 100\,\mathrm{GeV}\right)}{\mathrm{GeV}^{-1}\mathrm{cm}^{-2}}\right]  = 
	\left[ \frac{d}{10\,\mathrm{Mpc}}\right]^{\,2} \,\times\,
	\left\lbrace
	\begin{array}{lll}
		0.035&\,\,\,&\tau_e = 100\,\mathrm{s} \\ 
		0.058&\,\,\,&\tau_e = 1000\,\mathrm{s}\\ 
		0.17&\,\,\,&\tau_e = 10000\,\mathrm{s}\\
	\end{array}
	\right.
\label{EqFluxLimits}
\end{equation}
Each limit is only valid under the assumption that the entire neutrino signal is contained in the corresponding time window. In other words, SN 2008D could have emitted at most 19 (41, 122) times more neutrinos than assumed under the soft jet model with default parameters $\Gamma_b=3$ and $E_j=10^{51.5}\,\mathrm{erg}$. A higher flux would have been observed by IceCube with a probability of 90\%.

The primary systematic uncertainty in these limits stems from a possible bias in signal simulation, i.e. the value of $s$. Systematics for IceCube 22 have been studied by Abbasi et al. (\cite{pointsource}) and lead to a $\sim$15\% uncertainty in $s$, corresponding to a $\,^{+17}_{-13}$ percent shift in the limits. Incorporating the uncertainty of the BDT classification response, that is decreasing the signal prediction and increasing the background expectation by the corresponding uncertainty resulted in a negligible shift of $\sim$0.5\% in the limits. 

\begin{figure}
\centering
\caption{Spectrum of SN 2008D according to the soft jet model for different assumed jet Lorentz factors and under the assumption that the jet is pointing towards Earth.}
\vspace{2mm}
\includegraphics[width=\linewidth]{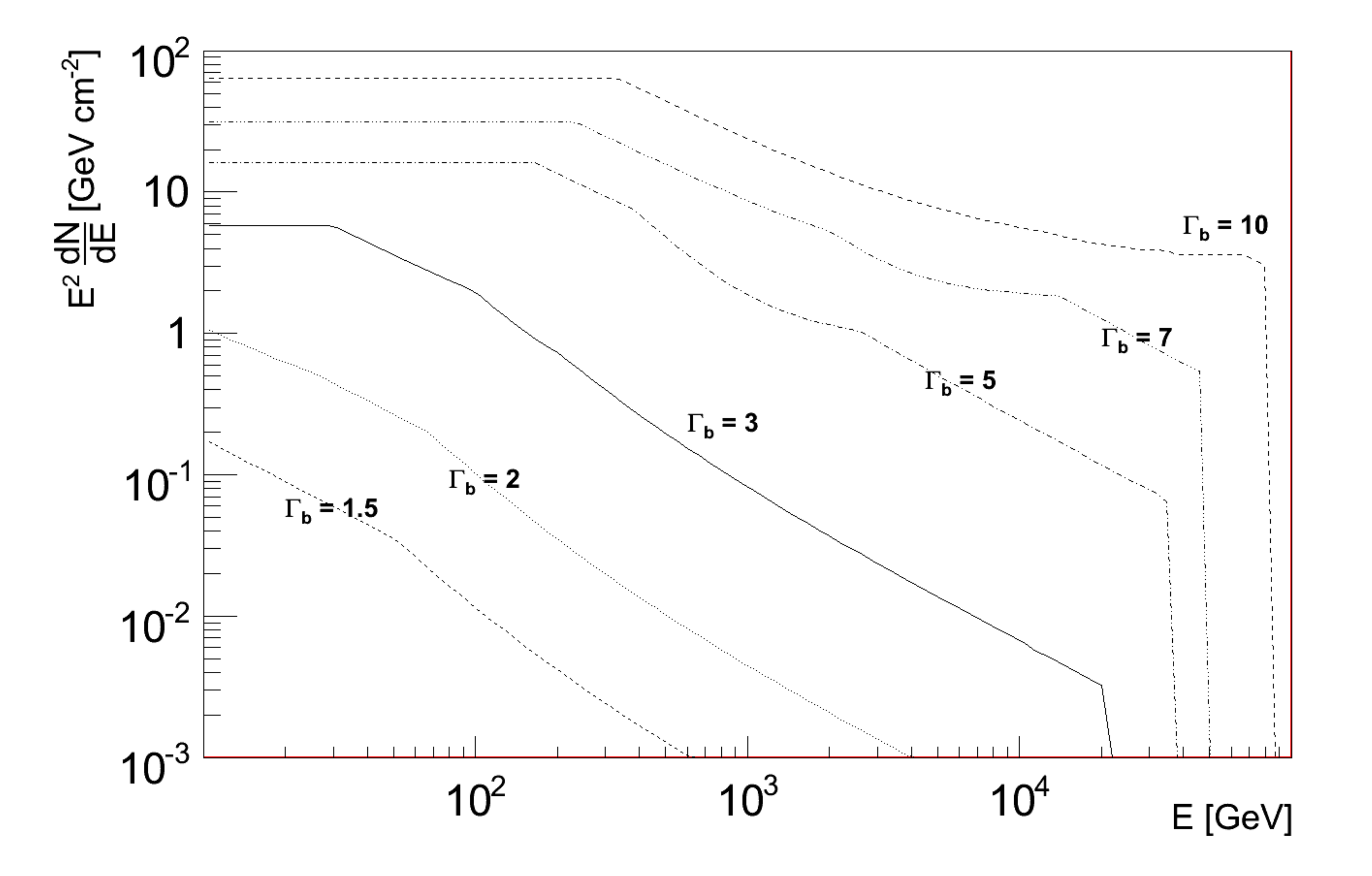}
\label{FigSpectrumOfGamma}
\end{figure}

Next, we wish to constrain the main parameters of the model, the kinetic energy release $E_j$ and the Lorentz factor of the jet $\Gamma_b$. 
Due to the significant $\Gamma_b$ dependence of the hadronic break energy $E_{\nu,cb}^{\pi/K\,(1)}\propto E_j^{-1}\,\Gamma_b^5$ and the radiative cooling break energy $E_{\nu,cb}^{\pi/K\,(2)}\propto \Gamma_b$, the number and spectral distribution of produced neutrinos depends strongly on $\Gamma_b$ (see Fig. \ref{FigSpectrumOfGamma}). Moreover, the flux is scaled with $E_j\,\Gamma_b^2$ which accounts for the energy release and the beaming of the neutrino emission. At high boost factors, radiative cooling of mesons sets in at lower energies than hadronic cooling, i.e. $E_{\nu,cb}^{\pi\,(1)}>E_{\nu,cb}^{\pi\,(2)}$ $\left(E_{\nu,cb}^{K\,(1)}>E_{\nu,cb}^{K\,(2)}\right)$ for $\Gamma_b \gtrsim 4$  $\left(\Gamma_b \gtrsim 9\right)$.

\begin{figure}
\centering
\caption{Expected number of events as a function of the assumed jet Lorentz factor $\Gamma_b$ under the assumption that the jet is pointing towards Earth. The plotted numbers correspond to a $10^{\circ}$-signal-region and cut level 3 at which the background rate is $0.03$ Hz. }
\vspace{2mm}
\includegraphics[width=\linewidth]{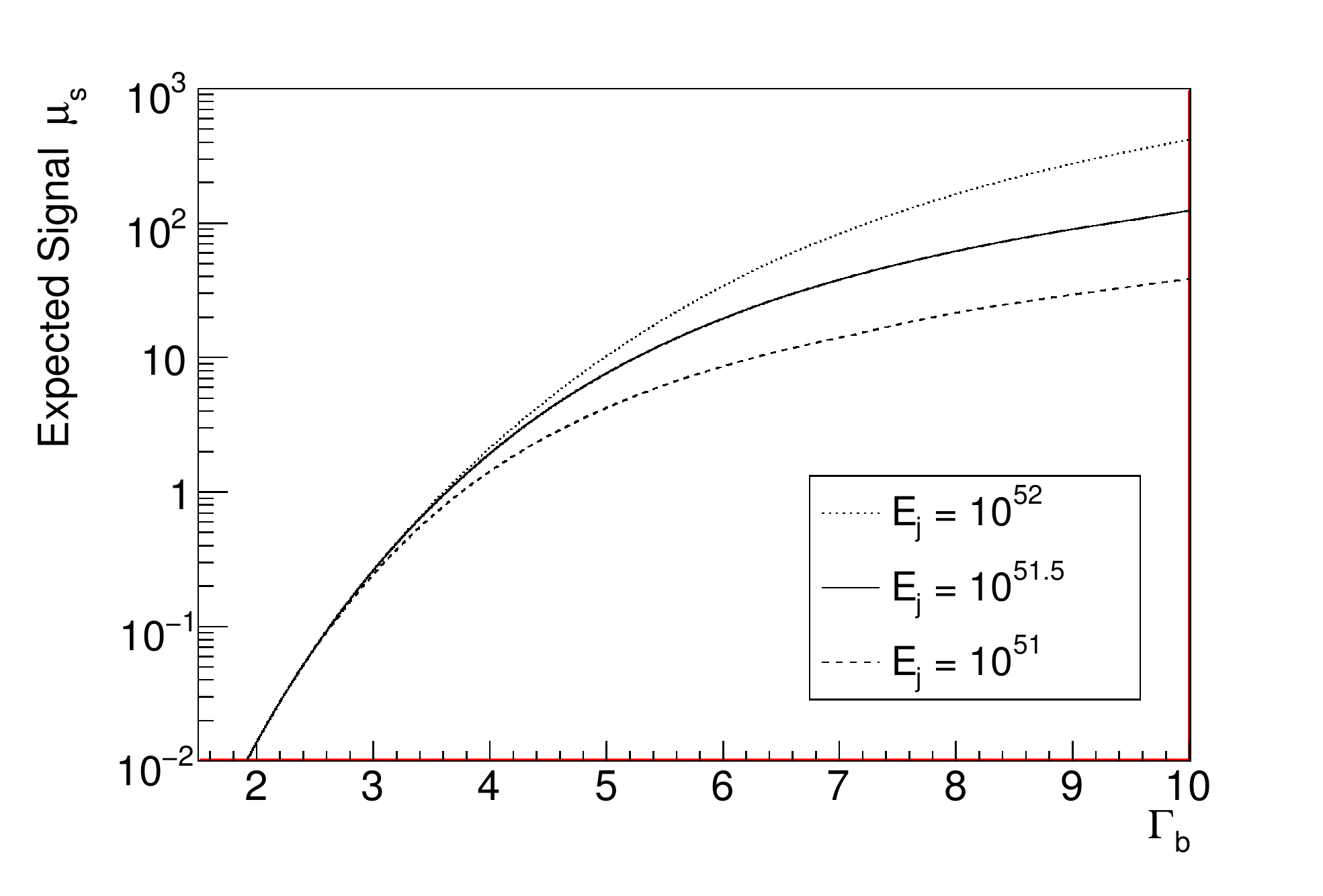}
\label{FigSignalOfGamma}
\end{figure}

\begin{figure}
\centering
\caption{Constraints on the jet parameters $E_j$ and $\Gamma_b$ where $E_{51.5}=10^{51.5}\,\mathrm{erg}$. For each assumed emission time scale $\tau_e$, the colored regions are ruled out at 90\% confidence level.}
\vspace{2mm}
\includegraphics[width=\linewidth]{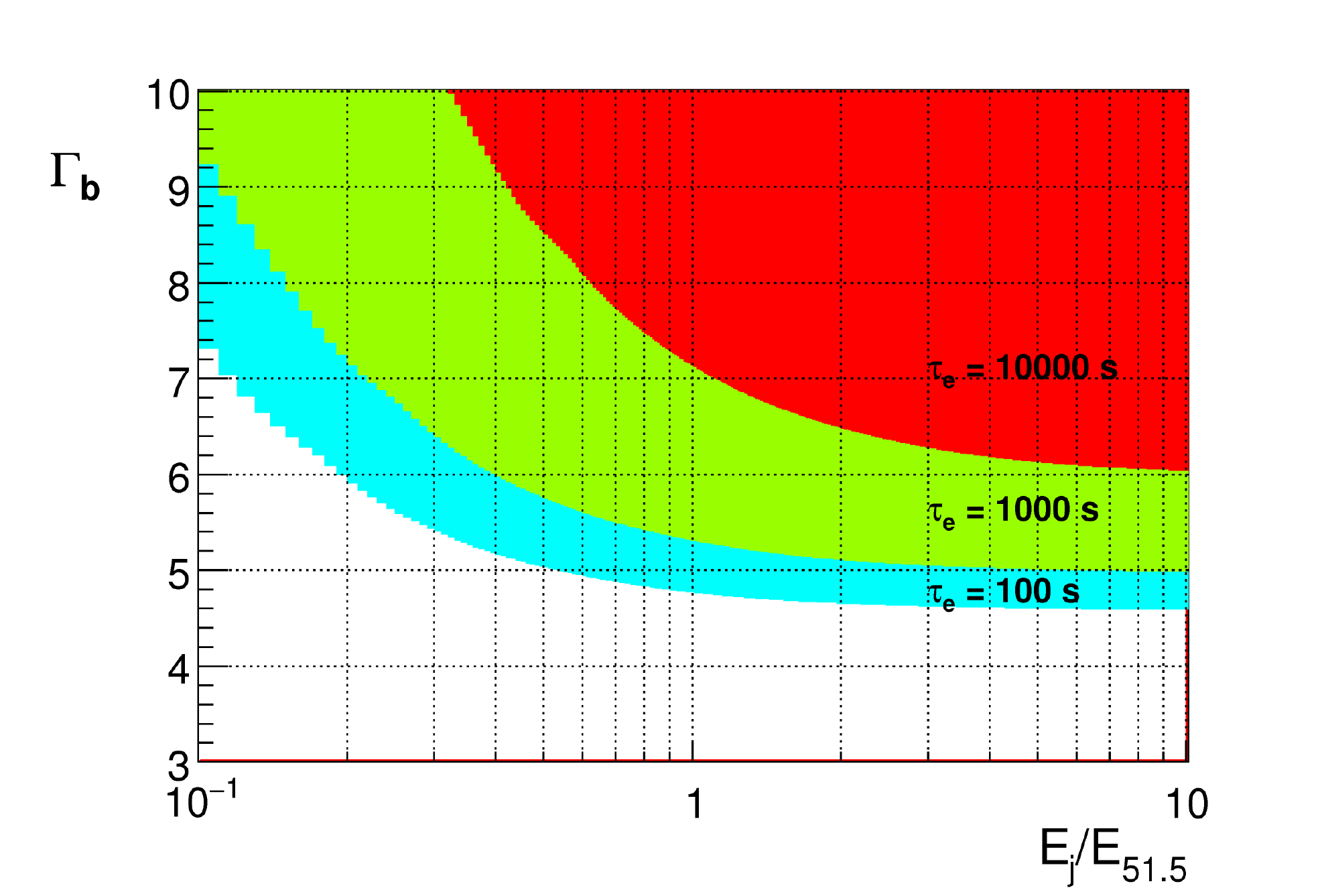}
\label{FigConstraints}
\end{figure}

To derive constraints on $\Gamma_b$ and $E_j$, we calculated the signal expectations in the intervals $\Gamma_b = 1.5\,-\,10$ and $E_j = 10^{51}\,-\,10^{52}\,\mathrm{erg}$. As Fig. \ref{FigSignalOfGamma} shows, the less efficient cooling as well as stronger beaming in more relativistic jets leads to a drastic increase in the signal expectation. Increasing $\Gamma_b$ places more neutrinos at high energies $\gtrsim$1 TeV where IceCube is more sensitive, though the corresponding reduction in the jet opening angle leads to smaller probability of jet detection. The measured signal upper limit $\bar{s}=2.44$ and the signal predictions $s_i\left(\Gamma_b,\,E_j\right)$ for each window can be used to constrain the jet parameters $E_j$ and $\Gamma_b$ through $s_i\left(\Gamma_b, E_j\right) < \bar{s}_i$. Values of $\Gamma_b$ and $E_j$ not fulfilling this relation are ruled out at 90\% CL. These limits are illustrated in Fig. \ref{FigConstraints}. 

Finally, the scenario proposed by Koers and Wijers (\cite{koers}) shall be examined briefly. Assuming that meson re-acceleration leads to a simple power law neutrino spectrum in the relevant energy range (roughly 100 GeV - 10 PeV) the source spectrum can be approximated by an $E^{-\gamma}$-law with a high-energy cutoff at $10\,\mathrm{PeV}$. For the three values of the spectral index $\gamma$ discussed by Koers and Wijers, this analysis yields the following upper limits for an assumed emission time scale of $\tau_e = 100\,\mathrm{s}$:
\begin{equation}
\frac{E^{\gamma}\bar{\Phi}^{(90)}}{\mathrm{GeV}^{\gamma - 1}\,\mathrm{cm}^{-2}} =
	\left\lbrace
	\begin{array}{lll}
		0.102&\,\,\,&\gamma = 2 \\ 
		1.62&\,\,\,&\gamma = 2.3\\ 
		21.9&\,\,\,&\gamma = 2.6\\
	\end{array}
	\right.
\label{EqSimpleLimits}
\end{equation}
For longer emission time scales, these limits scale as in (\ref{EqFluxLimits}).

\section{Summary and Outlook}

We have searched for high-energy muon neutrinos in coincidence with SN 2008D using data from the IceCube 22 string detector. Using a blind analysis optimized with experimental background and simulated signal data, we observed no events which passed the cuts. From the non-observation, we have derived first constraints on the soft jet model for core collapse SNe under the condition that the predicted jet was pointing in the direction of the Earth.

Given the strong dependence of the signal expectation on the model parameters, the non-detection of neutrinos places significant constraints on the principal model parameters. A two dimensional parameter scan in $\Gamma_{b}$ and $E_j$ shows that the jet Lorentz factor is generally constrained to $\Gamma_b<4$ for jet energies $E_{j}>10^{51}\,\mathrm{erg}$. As mentioned above, the constraints quoted here only hold if the assumed jet of SN 2008D was pointing towards Earth. 

IceCube is now operating in an additional mode, scanning online data for neutrino bursts, i.e. two nearly collinear neutrinos within 100 s, in real time. If a burst is detected, IceCube triggers optical follow-up observations searching for a SN in the corresponding direction (Franckowiak et al. \cite{ofu}). Constantly monitoring the entire northern sky, this approach has the potential to generalize the constraints obtained from studying individual objects.

\begin{acknowledgements}
We acknowledge the support from the following agencies:
U.S. National Science Foundation-Office of Polar
Program, U.S. National Science Foundation-Physics Division,
University of Wisconsin Alumni Research Foundation,
U.S. Department of Energy, and National Energy
Research Scientific Computing Center, the Louisiana
Optical Network Initiative (LONI) grid computing resources;
Swedish Research Council, Swedish Polar Research
Secretariat, and Knut and AliceWallenberg Foundation,
Sweden; German Ministry for Education and
Research (BMBF), Deutsche Forschungsgemeinschaft
(DFG), Germany; Fund for Scientific Research (FNRSFWO),
Flanders Institute to encourage scientific and
technological research in industry (IWT), Belgian Federal
Science Policy Office (Belspo); the Netherlands Organisation
for Scientific Research (NWO); M. Ribordy
acknowledges the support of the SNF (Switzerland);
A. Kappes and A. Groß acknowledge support by the EU
Marie Curie OIF Program; J. P. Rodrigues acknowledge
support by the Capes Foundation, Ministry of Education
of Brazil.
\end{acknowledgements}

\end{document}